\documentclass[preprint]{aastex}
\shortauthors{THORSTENSEN \& KIRKPATRICK}
\shorttitle{Serendipitous L-dwarf Parallax}
\begin{document}
\title{Serendipitous Discovery and Parallax of a Nearby L-Dwarf
\footnote{Based in part on
observations obtained at the Michigan-Dartmouth-MIT Observatory.}
}

\author{John R. Thorstensen}
\affil{Department of Physics and Astronomy\\
6127 Wilder Laboratory, Dartmouth College\\
Hanover, NH 03755-3528 \\
{\tt j.thorstensen@dartmouth.edu}}
\author{J. Davy Kirkpatrick}
\affil{Infrared Processing and Analysis Center, M/S 100-22\\
California Institute of Technology\\
Pasadena, CA, 91125 \\
{\tt davy@ipac.caltech.edu}}

\begin{abstract}
A field star serendipitously observed in a  
parallax program proved to have a proper motion of $562$ 
mas yr$^{-1}$ and a parallax of $82 \pm 2$ mas.
The star is identified with 2MASS J07003664+3157266.  A Keck
LRIS spectrum shows its spectral type to be L3.5, as expected
from its infrared and optical colors and absolute magnitude.
This object had not been previously recognized as 
an L dwarf, perhaps because of crowding at its relatively low
Galactic latitude ($b = +15.8$ degrees).
\end{abstract}

\section{Introduction}

The infrared sky surveys 2MASS, SDSS, 
and  DENIS \citep{skrut97,york00,epch94}
yielded a bountiful
harvest of very cool low-luminosity dwarf stars, which prompted
the extension of the conventional spectral classification sequence 
to $L$ and $T$ \citep{kir99}.  

An effort is underway to determine parallaxes for as many
$L$- and $T$-dwarfs as possible.  This is difficult, since these
objects can be faint in the optical band where most parallax
programs operate.  A modest number of precise parallaxes have been 
secured in the optical \citep{dahn02} and with infrared array detectors
\citep{tinney03,vrba02}, but more parallaxes are needed.  We describe
here the chance discovery of a nearby L dwarf, as a field star 
in a parallax sequence aimed at a different object.  Because of
its unusual mode of discovery, the object was `born' with a 
parallax.

\section{Observations}

\subsection{MDM Astrometry and Photometry}

One of us (JRT) is measuring parallaxes for a sample of cataclysmic
binaries using the 2.4m Hiltner telescope at MDM Observatory on Kitt
Peak, Arizona.  Technical details of this program will be given
elsewhere; a brief summary follows.  A SITe CCD is used with a
Kron-Cousins $I$-band filter to image an 8-arcminute square field at a
scale of 275 mas per pixel.  At each epoch a dozen or so short exposures
(25 s for the field considered here) are taken to avoid saturating
program or comparison stars and to average out the random centering
errors.  Image centers are measured using the IRAF implementation of
DAOPHOT \citep{stetsondao}.  A large number of stars are measured, not
only to constrain the plate solution adequately but also to empirically
estimate the scatter in the resulting parallaxes.  The scale and
orientation of each field is derived from fits to the USNO A2.0 catalog
\citep{mon96}.  An iterative procedure maps the star centers from each
picture onto a standard set of tangent-plane coordinates, and offsets 
as a function of time are derived, which are then fit for parallax and 
proper motion.  When conditions are appropriate, standardized photometry 
in $V$ and $I$ is obtained using \citet{landolt} standards for calibration.   

In order to verify the procedures used in the MDM parallax program and
to estimate the external accuracy, the fields of five `parallax
standards' were included.  These were selected from among the LHS stars
measured by \citet{monet92}.  The agreement is satisfactory, showing
that the procedures are essentially correct.  During the reductions it
was noticed that a star near the edge of the field of LHS 1889 showed a
large proper motion and parallax and a very red $V - I$ color.  This
proved to be 2MASS J07003664+3157266 (hereafter 2MASS 0700+3157);  Table 1
lists its parameters.   Fig.~1 gives a finding chart, and Fig.~2 shows
its path across the sky and its parallactic ellipse.  The colors and
absolute magnitude in Table 1 yielded a preliminary classification of L3.  

Because 2MASS 0700+3157 is much redder than the reference-frame stars, we
must be careful with differential color refraction (DCR);
\citet{monet92} discuss the problem.  The $I$-band filter used here is
narrower and redder than that used by \citet{monet92}, which ameliorates
the problem by a factor of about four.  Even so, we did apply a
correction, and confined most of the exposures to within $\pm 1$ hour of
meridian passage.  One set of exposures -- unfortunately the only set with
a large negative $X$ parallax factor -- was taken at hour angles $+1^{\rm
h} \le H \le +2^{\rm h}$.  Excluding these exposures changed  $\pi_{\rm rel}$
from 81 to 79 mas.  For the exposure farthest from the meridian,
the full amplitude of the DCR correction relative to the comparison stars
(typically $V-I = +1$) was 13 mas; allowing a generous 30
percent uncertainty in the DCR coefficient would shift the centroid in that
image by 4 mas.  This is as bad as it gets, so DCR evidently does not
affect the parallax too seriously.

\subsection{Keck Spectroscopy}

The L dwarf candidate was confirmed spectroscopically on 2003 Jan 02 UT
using the Low Resolution Imaging Spectrograph (LRIS; \citealt{oke96}) at
the 10m W.\ M.\ Keck-I Observatory on Mauna Kea, Hawaii.  Two
consecutive 300 s exposures were obtained.  A 400 lines/mm grating
blazed at 8500 \AA\ was used with a 1$\arcsec$ slit and 2048$\times$2048
CCD to produce 7-{\AA}-resolution spectra covering the range 6300 --
10100 \AA.  An OG570 filter eliminated second-order light.  The data
were reduced and calibrated using standard IRAF routines.  
Quartz-lamp flat-field exposures of the dome were used to normalize the
response of the detector, and stellar spectra were extracted using {\it
apextract}.  The wavelength calibration is from a NeAr arc lamp
exposures taken immediately after the 2MASS 0700+3157 observations, and
the flux calibration is from an observation of Hiltner 600 \citep{ham94}
taken immediately before the observations of the target. Fig.~3 shows
the reduced spectrum.  The data are not corrected for telluric
absorption, so atmospheric O$_2$ bands at 6867-7000, 7594-7685 \AA\ and
H$_2$O bands at 7186-7273, 8161-8282, $\sim$8950-9300, $\sim$9300-9650
\AA\ are evident.


The spectral type was assigned following the guidelines established in
\citet{kir99}.  The resulting values of the classification
ratios defined in that paper are CrH-a = 1.58(2), Rb-b/TiO-b =
1.21(3-4), and Cs-a/VO-b = 1.18(3-4), and the fit to the \ion{K}{1}
doublet indicates a subtype of (3). This gives a final spectral type of
L3.5.  Using the human eye alone to judge the type results in the same
spectral type of L3.5 because the spectrum of 2MASS 0700+3157 is
morphologically intermediate between the spectral standards 
2MASSW J1146345+223053 (L3) and 2MASSW J1155009+230706 (L4)
taken with the same 
instrumental setup.  The spectral type is almost exactly as predicted on
the basis of the colors and absolute magnitude.
The spectrum shows neither Li I $\lambda$6708 absorption, 
nor H$\alpha$ emission, to equivalent width limits of 0.3 and 0.2 \AA\ 
respectively.

\section{Discussion}

Our spectrum does not provide an accurate radial velocity, but the
transverse velocity is well-determined from the proper motion and
parallax (see Table 1).  The kinematics appear typical of a disk
population object.

There does not appear to be any physical association between the
L dwarf and the original target, LHS 1889.
The two stars' proper motions are similar but significantly different,
and the parallax of the L dwarf puts it significantly closer than LHS
1889, for which \citet{monet92} obtain $\pi_{\rm rel} = 52.8 \pm 0.9$
mas, and we find $\pi_{\rm rel} = 48.2 \pm 1.4$ mas.  

As noted earlier, the number of L-dwarfs with known parallax is still
modest.  The unusual channel through which this object was discovered
immediately yielded a parallax and placed it well within the 25 pc limit
of the \citet{glies91} catalog of nearby stars.  

This is among the closer and apparently brighter L dwarfs, yet it
escaped detection until now, probably because of the relatively
crowded field ($b = 15.8$ degrees).  L dwarfs have mostly been selected
as 2MASS or DENIS detections without optical counterparts, a criterion
which is difficult to apply to fast-moving objects in crowded fields.
SDSS, another prolific source of L-dwarf discoveries, is not
covering low latitudes.
Interestingly, \citet{salim03} recently discovered an apparently
even closer L dwarf at low latitude in their proper motion survey.
It is very likely that other undiscovered L dwarfs lurk 
nearby at low latitude.

{\it Acknowledgments.} JRT thanks the NSF for support through
AST 9987334, and especially thanks Cindy Taylor and Bill Fenton
for taking some of the parallax pictures.  The MDM staff cheerfully
performed many extra instrument changes over the years to facilitate
this project. 

This publication makes use of data products from the Two Micron All Sky Survey,
which is a joint project of the University of Massachusetts and the Infrared
Processing and Analysis Center/California Institute of Technology, funded by
the National Aeronautics and Space Administration and the National Science
Foundation.

Data presented herein were obtained at the W. M. Keck Observatory from
telescope time allocated to the National Aeronautics and Space Administration
through the agency's scientific partnership with the California Institute of
Technology and the University of California. The Observatory was made possible
by the generous financial support of the W. M. Keck Foundation. The authors
wish to recognize and acknowledge the very significant cultural role and
reverence that the summit of Mauna Kea has always had within the indigenous
Hawaiian community. We are most fortunate to have had the opportunity to
conduct observations from this mountain.  J.D.K. acknowledges support of the
Jet Propulsion Laboratory, California Institute of Technology, which is
operated under contract by the National Aeronautics and Space Administration.
J.D.K. would also like to express thanks for the assistance at Keck provided by
Patrick Lowrance, Joel Aycock, Paola Amico, and Barbara Schaefer.

\clearpage
\begin{deluxetable}{lll}
\tabletypesize{\footnotesize}
\tablewidth{5.0truein}
\tablecolumns{3}
\tablecaption{Parameters of 2MASS 0700+3157}
\tablehead{
\colhead{Parameter} &
\colhead{Value} &
\colhead{Unit or comment} \\
}
\startdata
$\alpha$ & $7^{\rm h}\, 00^{\rm m}\, 36^{\rm s}.66$ & ICRS; epoch 2000. \\
$\delta$ & $+31^{\circ}\, 57'\, 25''.8$ & ICRS; epoch 2000. \\
$\mu$ & $561.6 \pm 0.8$ & mas yr$^{-1}$ \\
P.A. & $166.64 \pm 0.08$ & degrees \\
$V$ & $21.68 \pm 0.03$ & (a) \\
$V - I$ & $5.26 \pm 0.03$ & (a) \\
$J$ & $12.923 \pm 0.023$ & 2MASS \\
$H$ & $11.947 \pm 0.016$ & 2MASS \\
$K_s$ & $11.317 \pm 0.023$ & 2MASS \\
$V - K_s$ & $10.36 \pm 0.04$ & \\
$\pi_{\rm rel}$ & $81 \pm 2$ & mas \\
$\pi_{\rm abs}$ & $82 \pm 2$ & mas (b)\\
$1/\pi_{\rm abs}$ & $12.2 \pm 0.3$ & pc \\
$m - M$ & $+0.48 \pm 0.06$ & mag \\
$M_V$ & $21.20 \pm 0.07$ & mag \\
Sp. & L3.5 & \\
$v_T$ & $32.4 \pm 0.8$ & km s$^{-1}$ \\
$v_T$ (LSR) & 21 & km s$^{-1}$; (c) \\

\enddata
\tablenotetext{a}{From three $VI_{\rm KC}$ image pairs obtained 2003 Jan 29 UT.
The color terms in the standardization were small, but the object is much
redder than any of the standard stars, so systematic uncertainties are larger
than the formal errors shown.}
\tablenotetext{b}{The correction to absolute parallax is from color-based 
distance estimates for the stars used to define the astrometric grid.}
\tablenotetext{c}{Magnitude of the transverse velocity {\it only}, 
referred to the transverse components of the local standard of rest.}
\end{deluxetable}

\clearpage

\begin{figure}   
\caption{[Included as a jpg file for the astro-ph version.]  
Chart showing 2MASS 0700+3751 and LHS1889, the original
parallax target, from an MDM 2.4m I-band image.  The lines 
indicate the proper motion for the next 50 years.  The proper
motions are similar, but significantly different.
}
\end{figure}

\begin{figure}   
\plotone{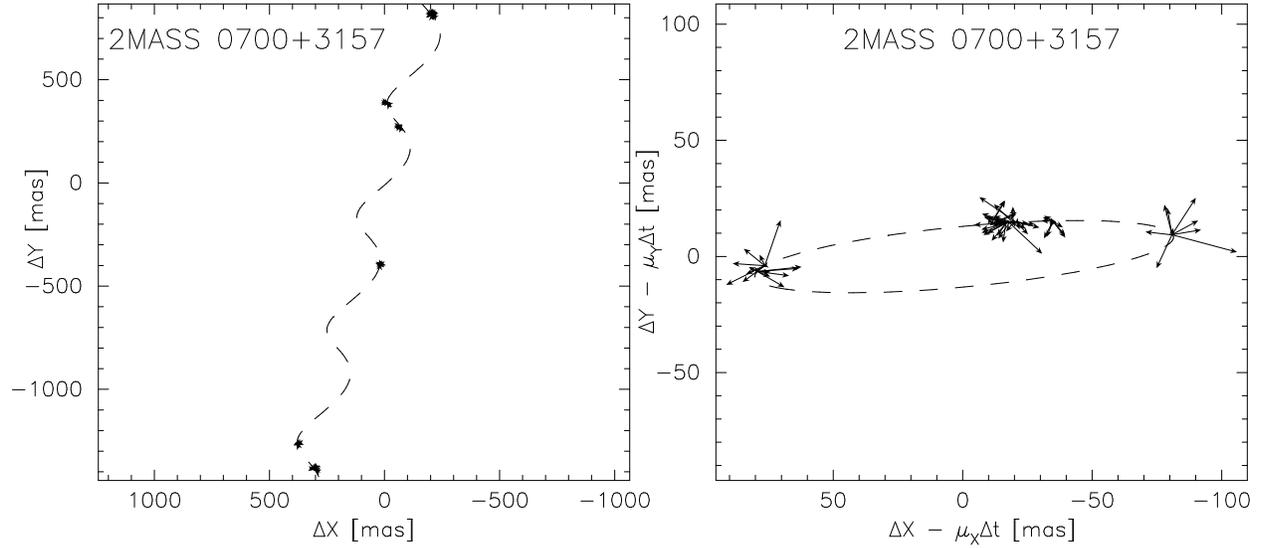}
\caption{{\it Left panel:} Positions of 2MASS 0700+3157,
referred to its nominal position.  The dashed curve is the 
trajectory computed for its fitted parallax and proper motion.
{\it Right panel:} The same data, but in the reference frame
moving with the computed proper motion.  The tip of each
arrow is the observation from a single image, and the tail
is the position computed for the epoch of that image from
the fitted parallax and proper motion.
}
\end{figure}

\begin{figure}   
\epsscale{0.85}
\plotone{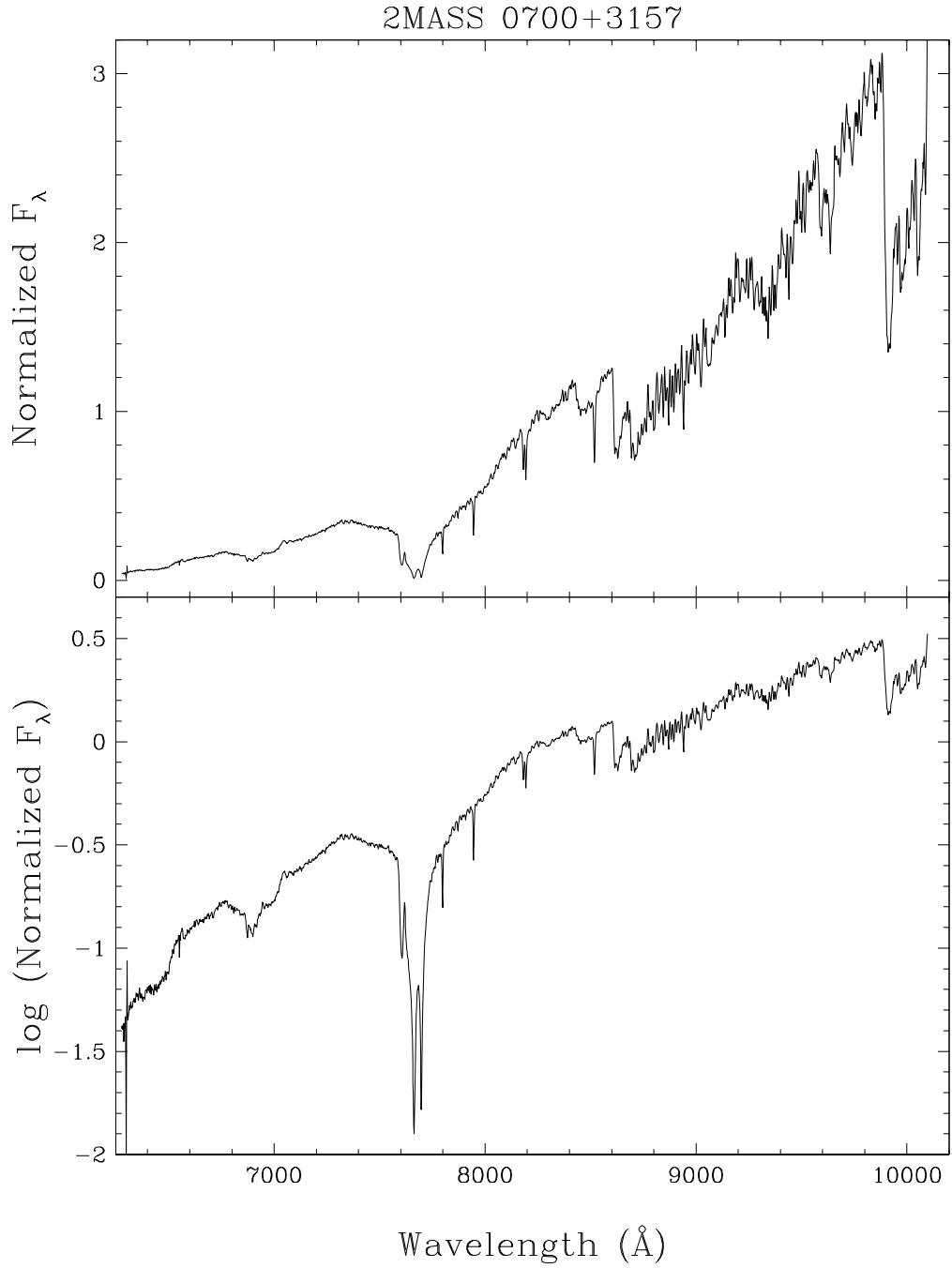}
\caption{Keck LRIS spectrum of 2MASS 0700+3751, displayed
in linear and logarithmic flux units.
}
\end{figure}

\end{document}